\documentclass[12pt]{article}

\usepackage{physics} 
\usepackage{siunitx} 
\usepackage{enumerate} 
\usepackage{graphicx}

\usepackage{amssymb}

\usepackage{cite}

\begin{document}

\title{Average trapping time on a type of horizontally segmented 3 dimensional Sierpinski gasket network with two types of locally self-similar structures}
\author{Zhizhuo Zhang; Bo Wu}
\date{\today}

\maketitle

\begin{abstract}
As a classic self-similar network model, Sierpinski gasket network has been used many times to study the characteristics of self-similar structure and its influence on the dynamic properties of the network. However, the network models studied in these problems only contain a single self-similar structure, which is inconsistent with the structural characteristics of the actual network models. In this paper, a type of horizontally segmented 3 dimensional Sierpinski gasket network is constructed, whose main feature is that it contains the locally self-similar structures of the 2 dimensional Sierpinski gasket network and the 3 dimensional Sierpinski gasket network at the same time, and the scale transformation between the two kinds of self-similar structures can be controlled by adjusting the crosscutting coefficient. The analytical expression of the average trapping time on the network model is solved, which used to analyze the effect of two types of self-similar structures on the properties of random walks. Finally, we conclude that the dominant self-similar structure will exert a greater influence on the random walk process on the network.
\end{abstract}

Keywords: Sierpinski Gasket; Random Walks; Self-Similarity; Average trapping time.

\section{Introduction}
In recent years, with the deepening of research on the properties of complex network structures, network models with specific global characteristics such as scale-free networks\cite{barabasi2003scale} and small-world\cite{watts1998collective} networks have gradually become research hotspots\cite{newman2018networks}.
Among these special global structural characteristics, the global self-similar structure\cite{song2005self} of the network has been proved to be an important topological characteristic of many real networks, and affects many dynamic behaviors on this network\cite{malozemov2003self}.
The Sierpinski gasket network, as a classic self-similar network model, has been paid attention to by many scholars\cite{le2015complex,wang2017average,jiang2018some}. 
Therefore, many topological and dynamic properties on the Sierpinski gasket network have been systematically studied, including:  hamiltonicity\cite{xue2012hamiltonicity}, Laplacian spectrum\cite{bentz2010analytic,rammal1984spectrum}, unbiased random walk\cite{kozak2002analytic}, the hub number\cite{lin2011hub}, spanning trees\cite{chang2007spanning}, independent spanning trees\cite{hasunuma2015structural}, the outer connected domination number\cite{chang2016outer}, etc.
In addition, the Sierpinski gasket network not only plays an important role in the theoretical study of the properties of self-similar structures, but also plays a key role in certain application fields, such as: fractal polymeric networks\cite{rothemund2004algorithmic,shang2015assembling} and  molecular self-assembly\cite{jurjiu2002strange,newkome2006nanoassembly}, etc.

In fact, how the random walk property on the network, as a basic dynamic property, is affected by the self-similar structure has always been a hot issue of concern. 
For example, on the Sierpinski gasket network, R.A Guyer et al. studied the diffusion problem\cite{guyer1984diffusion}; B Meyer studied the problem of mean time to absorption\cite{meyer2012exact}; Zhang et al. studied the problem about hitting time\cite{qi2020hitting}; C.P Haynes et al. studied the problem of global first-passage times of fractal networks\cite{haynes2008global}. 
Among the related properties of random walks, the average trapping time (ATT) has strong representativeness and application prospects\cite{montroll1969random,brodbeck2000new}, so it has also attracted the attention of our team. 
The ATT problem on the 3-level Sierpinski gasket network\cite{wu2020average} and the joint Sierpinski gasket network\cite{zhang2021mean} is studied, and a series of conclusions are obtained, including the calculation method of ATT analytical expression for high-level Sierpinski gasket networks and the effect of local self-similar structure on random walk. However, the above-mentioned problems all focus on the network model with a single self-similar structure, and the self-similar structure of the real network is often more complex and diverse.

Thereupon, a natural question is when the network has two different self-similar structures at the same time, what rules will the random walk property show, and whether there is an analytical expression of ATT and then give an exact conclusion through numerical analysis? 
In order to answer the above questions, a kind of horizontally segmented 3 dimensional Sierpinski gasket network is constructed in this paper, which possesses both the local self-similar structure characteristics of the 3 dimensional Sierpinski gasket network and the 2-dimensional Sierpinski gasket network, and by adjusting the segmentation coefficient, the dominant self-similar structure of the network will also change.
The goal of this paper is to solve the analytical expressions of ATT on the number of iteration and segmentation coefficient on the network, and then analyze the effect of the change of the self-similar structure on the properties of random walks.

The main content of the paper is divided into the following 5 parts: In the second section, the construction method and structural properties of the related Sierpinski gasket network are introduced, including 2 dimensional Sierpinski gasket network, 3 dimensional Sierpinski gasket network, horizontally segmented 3 dimensional Sierpinski gasket network and a type of auxiliary network; In the third and fourth sections, the analytical expressions of the ATT of the 3 dimensional Sierpinski gasket network and the auxiliary network are solved separately; in the fifth section, the analytical expression of the ATT of the horizontally segmented 3 dimensional Sierpinski gasket network is obtained and the numerical correlation between ATT and the segmentation coefficient is analyzed; Finally, in the sixth section, the content and conclusions of the paper are summarized.

\section{The construction method of networks}

In this section, we first introduce the iterative construction method of the 3 dimensional Sierpinski gasket network. Then, the segmentation variable $s$ is introduced into the network to construct a class of horizontally segmented 3 dimensional Sierpinski gasket network.

\begin{figure}[t]
\centering
\includegraphics[scale=0.45]{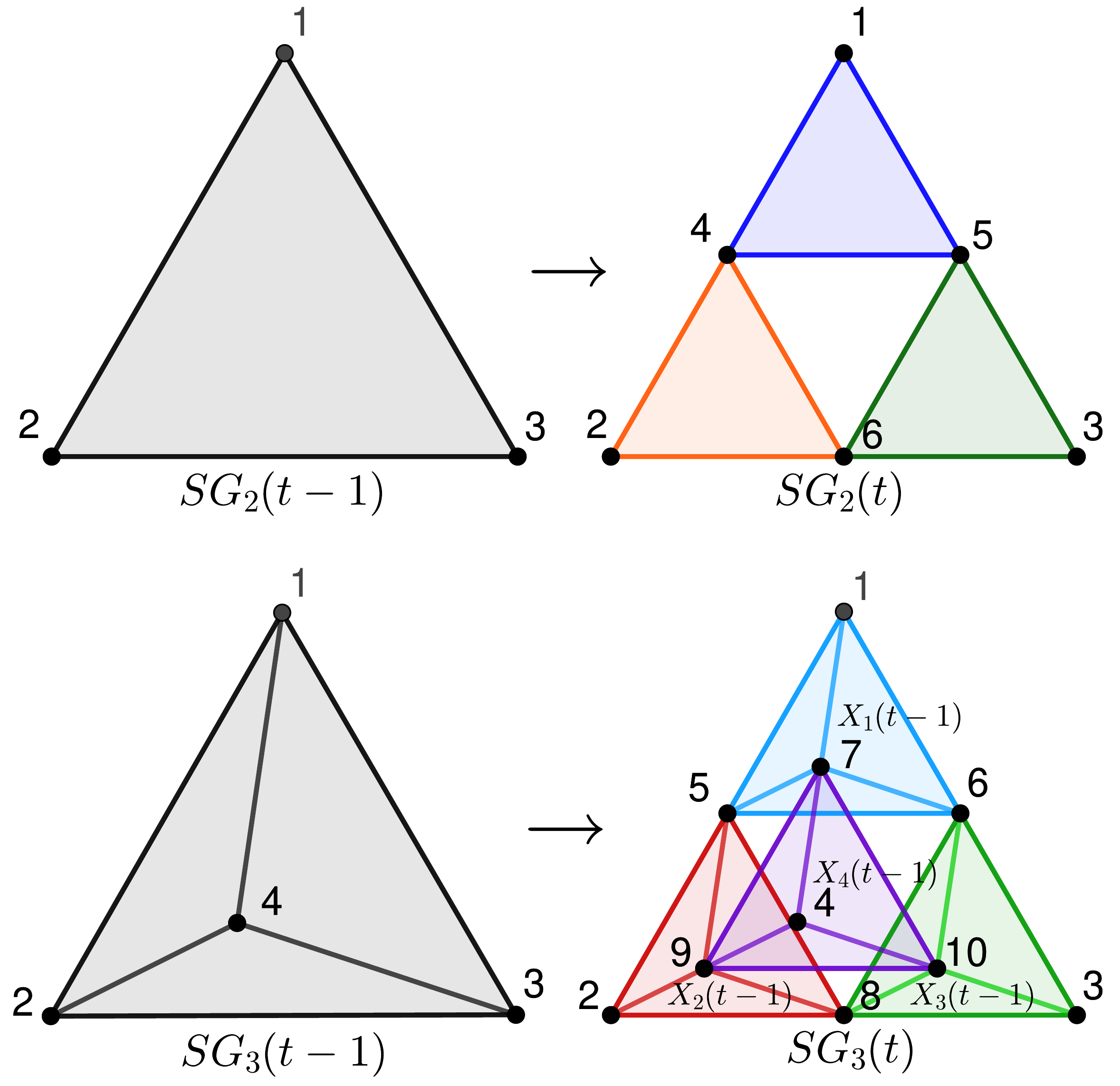}\label{fig1}
\caption{The Structures of $SG_2(t)$ and $SG_3(t)$}
\end{figure}

\subsection{The structure of the 3 dimensional Sierpinski gasket network}

Similar to the classic 2 dimensional Sierpinski gasket network, the 3 dimensional Sierpinski gasket network is also constructed by a iterative method. First, the number of iterations of the network is denoted as $t$, and the 2 dimensional and 3 dimensional Sierpinski gasket networks are denoted as $SG_2(t)$ and $SG_3(t)$, respectively. 
Then, as the initial network of $SG_2(t)$, $SG_2(0)$ is a complete graph (there is an edge between any two nodes) composed of three nodes. Similarly, as the initial network of $SG_3(t)$, $SG_3(0)$ is a complete graph composed of four nodes.
The iterative construction method of network $SG_2(t)$ is similar to $SG_3(t)$, so we will mainly introduce the structural characteristics of $SG_3(t)$ next.

As shown in the lower half of Fig(1), the network $SG_3(t)$ can be obtained by splicing four previous generation networks $SG_3(t-1)$, where the structure of the local region $X_i(t-1)$ $(1\le i \le 4)$ inside the network $SG_3(t)$ is equivalent to the structure of the $SG_3(t-1)$. Based on the structure of the 3-dimensional network $SG_3(t)$, it is naturally easy to obtain the structure model of the 2-dimensional network $SG_2(t)$, as shown in the upper part of Fig(1), and we will not repeat it here.

In fact, it can be found that the network $SG_3(t)$ is constructed by $4^{t-k}$ local regions $X_i(k)$. Therefore, the iterative process of the network $SG_3(t)$ can be regarded as replacing all the smallest local structures $X_i(0)$ in the network $SG_3(t-k)$ with $X_i(k)$, where $1\le k\le t-1$. The equivalence between the different local structures and the similarity between the local structure and the overall structure are just the embodiment of the global self-similar structure. In order to facilitate subsequent node labeling and classification, we specify the iterative method of $k=1$ as the standard iterative process.

Then, the nodes in $SG_3(t)$ are marked according to the order of generation, in which the nodes on the initial network $SG_3(0)$ are denoted as $1$, $2$, $3$, and $4$ respectively; the $6$ nodes generated after one iteration are denoted as $5$, $6$, $7$, $8$, $9$ and $10$, as shown in Fig(1). Furthermore, in the network $SG_3(t)$ the set of local regions $X_i(k)$ $(0\le k\le t)$ is denoted as $\Delta^k(t)$, the set of all nodes is denoted as $V(t)$, and the set of all edges is denoted as $E(t)$.
In addition, the symbol $|\cdot|$ is used to indicate the number of elements in the set. Therefore, we can easily obtain:
\begin{align}
    |\Delta^k(t)|&=4^{t-k},\label{2.1}\\
    |E(t)|&=6\cdot|\Delta^0(t)|=6\cdot 4^t,\label{2.2}\\
    |V(t)|&=|V(0)|+6\cdot\sum_{i=0}^{t-1}|\Delta^0(i)|=2\cdot 4^t+2.\label{2.3}
\end{align}
Since the local region $X^i(k)$ $(1\le i\le 4^{t-k})$ in the network $SG_3(t)$ is equivalent to the network $SG_3(k)$, we denote the nodes in $X^i(k)$ corresponding to nodes $1$, $2$, $3$ and $4$ in the $SG_3(k)$ as $1_i$, $2_i$, $3_i$ and $4_i$. The set of all nodes in $X^i(k)$ is denoted as $V_i^k(t)$, $\bar{V}_i^k(t)=\{1_i, 2_i, 3_i, 4_i\}$ and $\hat{V}_i^k(t)=V_i^k(t)/\bar{V}_i^k(t)$. Therefore, we have
\begin{align}
    V(t)=\bigcup_{i=1}^{|\Delta^k(t)|}(\bar{V}_i^k(t)\cup \hat{V}_i^k(t))=\bar{V}^k(t)\cup \hat{V}^k(t),
\end{align}
where $\bar{V}^k(t)=\bigcup_{i=1}^{|\Delta^k(t)|}\bar{V}_i^k(t)$, $\hat{V}^k(t)=\bigcup_{i=1}^{|\Delta^k(t)|}\hat{V}_i^k(t)$ and $\hat{V}_i^k(t)$ is a disjoint set, but $\bar{V}_i^k(t)$ is not. Similarly, the three nodes in $X^i(k)$ corresponding to $5$, $6$, and $7$ in $SG_3(k)$ are denoted as $5_i$, $6_i$, and $7_i$, respectively.

\subsection{The structure of the horizontally segmented 3 dimensional Sierpinski gasket network}

\begin{figure}[t]
\centering
\includegraphics[scale=0.25]{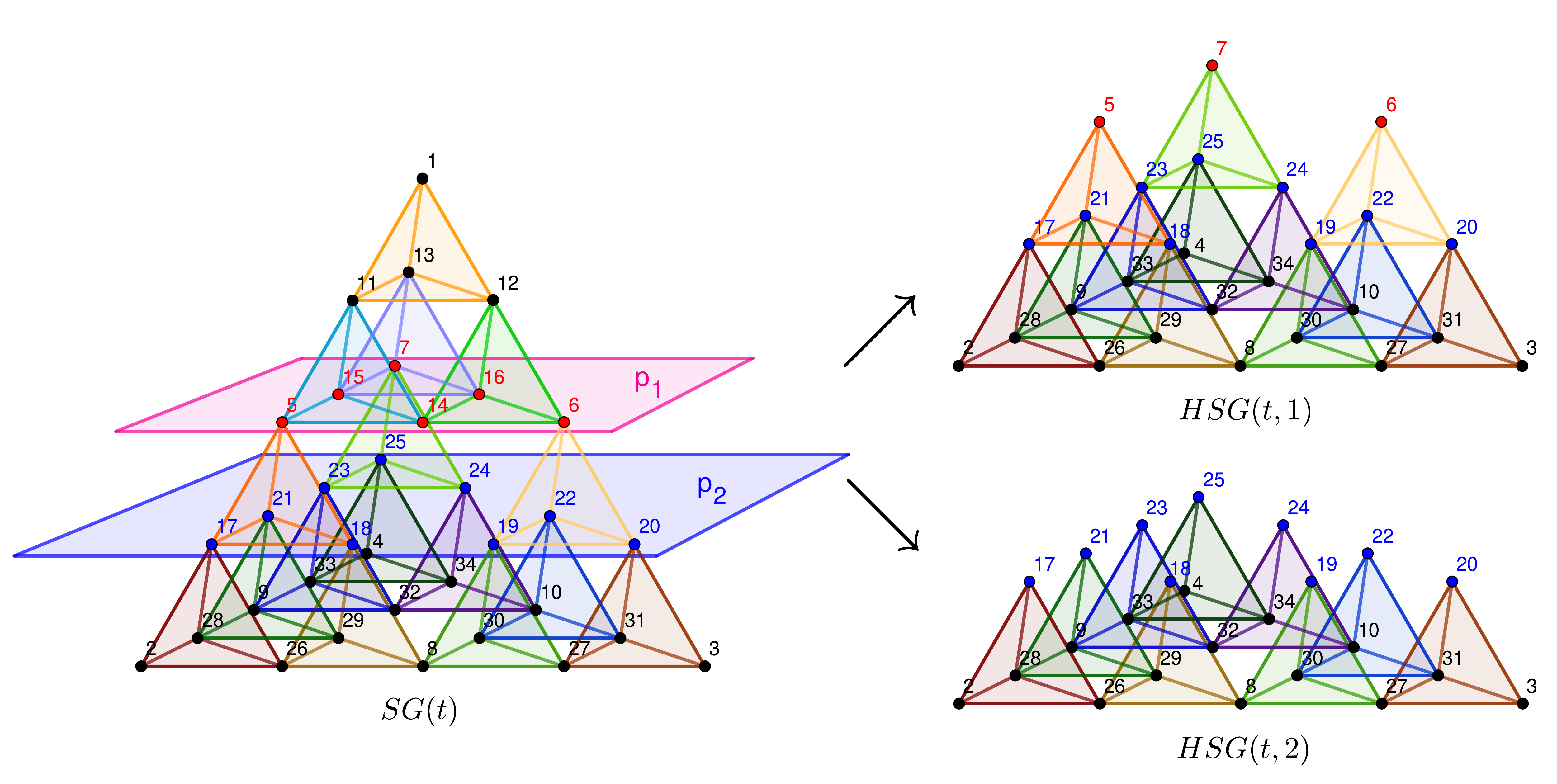}\label{fig2}
\caption{The Structures of $HSG(t,1)$ and $HSG(t,2)$}
\end{figure}

Based on the $3$ dimensional Sierpinski gasket network, a class of horizontally segmented $3$ dimensional Sierpinski gasket network model, denoted as $HSG(t,s)$, can be constructed next. 

First, the segmentation plane $p_1$ is defined as the plane determined by nodes $5$, $6$, and $7$ in the 3 dimensional space where $SG_3(t)$ is located. The network $SG_3(t)$ is divided into two parts by the plane $p_1$, and then the upper half is discarded and only the lower half is retained, in which all edges in the plane $p_1$ are discarded. After discarding the edges, the isolated nodes in the plane $p_1$ are discarded, and the remaining nodes are retained. So far, the network $HSG(t,1)$ is constructed, which is composed of three local regions $X_i(t-1)$, as shown in the upper part of Fig(2), where, the red nodes are all located on the plane $p_1$, but only three nodes $5$, $6$, and $7$ are reserved. 
Then, in any region $X_i(t-1)$ in the $HSG(t,1)$, the plane determined by the nodes $5_i$, $6_i$, and $7_i$ is defined as the segmentation plane $p_2$. Similarly, after being split along the plane $p_2$, the lower half of the network is defined as $HSG(t,2)$. It is worth noting that the $HSG(t,2)$ is formed by splicing 9 local regions $X_i(t-2)$, as shown in the lower part of Fig(2). From this, we can naturally deduce the position of the segmentation plane $p_s$ $(1\le s\le t)$ and the horizontally segmented $3$ dimensional Sierpinski gasket network $HSG(t,s)$ obtained after the network $SG_3(t)$ is divided along the plane, which is composed of $3^s$ local regions $X_i(t-s)$.

Obviously, the network $HSG(t,s)$ cannot be considered to have a global self-similar structure, because its local region $X_i(t-s)$ is not similar to the network as a whole. However, these local regions $X_i(t-s)$ themselves have self-similar structures, and these regions are equivalent. Therefore, this structural feature is called a local self-similar structure. In fact, this special segmentation method guarantees the parameterization of the segmentation plane, that is, the degree of network damage is described by the segmentation coefficient $s$ and the segmentation coefficient $s$ further determines the scale of the local self-similar structure in the network $HSG(t,s)$. 
These features also ensure that the property of random walk on $HSG(t,s)$ will necessarily be closely related to the segmentation coefficient $s$, and the influence of the change process of the network's self-similar structure on the property of random walk can be obtained by analyzing the change of $s$.

It is worth noting that the network $HSG(t,s)$ can actually be constructed by 2 dimensional and 3 dimensional Sierpinski gasket networks. In fact, $HSG(t,s)$ is established by replacing each of the smallest triangles in $SG_2(s)$ with a 3 dimensional Sierpinski network $SG_3(t-s)$, in which the three nodes $2$, $3$, and $4$ of $SG_3(t-s)$ correspond to the three nodes in the replaced smallest triangle. In order to facilitate subsequent calculations, we define the auxiliary network $ASG(x,y)$ here as the network model obtained by replacing each smallest triangle in the network $SG_2(y)$ with $SG_3(x)$ according to the above process. Therefore, we have $HSG(t,s)=ASG(t-s,s)$. Because of this, when the number of iterations $t$ is fixed, as the segmentation coefficient $s$ increases, the scale of the local self-similar structure $X_i(t-s)$ continues to decrease, but the characteristics of the 2-dimensional self-similar structure does increase accordingly.

Let the sets of nodes in the networks $HSG(t,s)$ and $ASG(x,y)$ be $V^{H}(t,s)$ and $V^{A}(x,y)$, respectively; the sets of edges are $E^{H}(t,s)$ and $E^{A}(x,y)$, respectively. It is easy to prove that the following relationships are established:
\begin{align}
    &|E^{A}(x,y)|=6\cdot 3^{y}\cdot 4^{x},~
    |V^{A}(x,y)|=2\cdot 3^{y} \cdot 4^{x} + \frac{1}{2}3^{y}+ \frac{3}{2};\\
    &|E^{H}(t,s)|=6\cdot 3^{s}\cdot 4^{t-s},~
    |V^{H}(t,s)|=2\cdot 3^{s} \cdot 4^{t-s} + \frac{1}{2}3^{s}+ \frac{3}{2}.
\end{align}

In addition, since $ASG(x,y)$ is composed of $3^y$ local self-similar regions $X^i(x)$ $(1\leq i \leq 3^y)$, the set of all nodes in region $X^i(x)$ is denoted as $V^{A}_i(x,y)$.
Let $\tilde{V}^{A}_i(x,y)=\{2_i,3_i,4_i\}$. Then, we can know that the set of all nodes in the network $SG_2(x)$, denoted as $\tilde{V}^{A}(x,y)$, satisfies:
\begin{align*}
    \tilde{V}^{A}(x,y)=\bigcup_{i=1}^{3^x}\tilde{V}^{A}_i(x,y).
\end{align*}
For the convenience of description, the three outermost nodes in $\tilde{V}^{A}(x,y)$ are denoted as $\alpha_2$, $\alpha_3$, and $\alpha_4$, which correspond to nodes $2$, $3$, and $4$ in the $HSG(t,s)$.

Consequently, let 
$$
\dot{V}^{A}_{i}(x,y)=V^{A}_i(x,y) /\tilde{V}^{A}_i(x,y)~~~ \textrm{and}~~~ \dot{V}^{A}(x,y)=\bigcup_{i=1}^{3^x}\dot{V}^{A}_i(x,y).
$$
Then, we have
$$
{V}^{A}(x,y)=\tilde{V}^{A}(x,y)\cup \dot{V}^{A}(x,y) ~~~ \textrm{and}~~~ \tilde{V}^{A}(x,y)\cap \dot{V}^{A}(x,y) = \varnothing
$$

Based on the structural relationship between the network $HSG(t,s)$ and the auxiliary network $ASG(x,y)$, if the ATT on the network $ASG(x,y)$ is solved, the ATT on the $HSG(t,s)$ is also obtained.

\section{ATT on $SG_3(t)$ }

In this section, we will discuss the ATT of trap node $2$ on network $SG_3(t)$. Before starting the formal calculation, the mark of the random walk on the network needs to be given. First of all, for any two nodes $u$ and $v$ on the network $SG_3(t)$, the average value of the first passage time of the walker from node $u$ to the node $v$ is defined as the mean first passage time (MFPT), denoted as $T_{u,v}^3(t)$. Here, node $v$ is called a target node. It is worth noting that the target node is not necessarily a single node, but can also be a node set $A$. At this time, $T_{u,A}^3(t)$ represents the MFPT for the walker to depart from node $u$ and arrive at any node in $A$ for the first time. 
In the network $SG_3(t)$, node $2$ is defined as a trap node, and then the mean capture time (MCT) of node $u$, denoted as $T_{u}^{3}(t)$, is defined as the average value of the time when the walker first arrives at the trap node from node $u$. Therefore, the MCT of node $u$ satisfy that $T_{u}^{3}(t)=T_{u,2}^{3}(t)$. In addition, we also stipulate $T_{2}^{3}(t)=0$. Next, the sum of the MCTs of all nodes on the network $SG_3(t)$, denoted as $T_{sum}^{3}(t)$, is defined as
\begin{align*}
    T_{sum}^{3}(t)=\sum_{u\in V(t)}T_{u}^{3}(t)=\sum_{u\in V(t)/\{2\}}T_{u,2}^{3}(t).
\end{align*}
Therefore, the ATT on $SG_3(t)$ with trap node $2$, denoted as $\langle T^3(t) \rangle$, is defined as 
\begin{align*}
    \langle T^3(t) \rangle=\frac{T_{sum}^{3}(t)}{|N(t)|-1}
    =\frac{1}{2\cdot4^{t}+1}\sum_{u\in V(t)}T_{u}^{3}(t).
\end{align*}

In the same way, the definition of ATT on the network $SG_2(t)$ can naturally be obtained. Of course, the ATT on the network $SG_2(t)$ has been solved, and the analytical expression for the sum of MCTs across all nodes, denoted as $T_{sum}^{2}(t)$, is as follows\cite{meyer2012exact,zhang2021mean}:
\begin{align}
    T_{sum}^{2}(t)=\frac{5}{2}\cdot3^{t}\cdot5^{t}+2\cdot5^{t}+\frac{1}{2}\cdot3^{t}.
\end{align}

In order to solve the ATT on the network $SG_3(t)$ with the trap node $2$, we first consider the random walk property on the first generation network $SG_3(1)$. Let the target node set $A=\{2,3,4\}$, $B=\{1,2,3,4\}$. From the random walk properties on the network and the structural symmetry of $SG3 (1)$, the following equations can be easily obtained:
$$
\left\{
\begin{array}{ll}
     T_{1,A}^{3}(1)=&\frac{1}{3}T_{5,A}^{3}+\frac{1}{3}T_{6,A}^{3}(1)+\frac{1}{3}T_{7,A}^{3}(1)+1 \\
     T_{5,A}^{3}(1)=&\frac{1}{6}T_{1,A}^{3} + +\frac{1}{6}T_{6,A}^{3} + \frac{1}{6}T_{7,6}^{3} +  \frac{1}{6}T_{8,A}^{3} +\frac{1}{6}T_{9,A}^{3} + 1\\
     T_{8,A}^{3}(1)=& \frac{1}{6}T_{5,A}^{3} + \frac{1}{6}T_{6,A}^{3} + \frac{1}{6}T_{9,A}^{3} + \frac{1}{6}T_{10,A}^{3} + 1\\
     T_{5,A}^{3}(1)=& T_{6,A}^{3}(1) = T_{7,A}^{3}(1)\\
     T_{8,A}^{3}(1)=& T_{9,A}^{3}(1) = T_{10,A}^{3}(1)
\end{array}
\right.
$$
It is easy to solve from the above group: 
$$
\lambda\triangleq T_{1,A}^{3}(1) = 6,~~ T_{5,A}^{3}(1)=5
~~ T_{8,A}^{3}(1)=4.
$$
Similarly, when $B$ is the set of target nodes, the following equations hold:
$$
\left\{
\begin{array}{ll}
     T_{5,B}^{3}(1)&=\frac{1}{6}T_{6,B}^{3}(1) + \frac{1}{6}T_{7,B}^{3}(1) + \frac{1}{6}T_{8,B}^{3}(1) + \frac{1}{6}T_{9,B}^{3}(1) + 1\\
     T_{5,B}^{3}(1) &= T_{6,B}^{3}(1) = T_{7,B}^{3}(1) = T_{8,B}^{3}(1) = T_{9,B}^{3}(1) = T_{10,B}^{3}(1)
\end{array}
\right.
$$
Therefore, it can be solved that
$$
\sigma \triangleq T_{5,B}^{3}(1) =3.
$$

As described by the global self-similarity of the network $SG_3(t)$, the structure of each local region $X_i(1)$ of the network is equivalent to $SG_3(1)$. For any $u\in \bar{V}_{i}^{1}(t)$, the node belongs to at most two local areas, denoted as $X_{i}(1)$ and $X_{j}(1)$, at the same time, and $X_{i}(1)$ and $X_{j}(1)$ are symmetrical with respect to the node $u$. Therefore, the walker starts from the node $u$, and after a random walk of the average time $\lambda$, it will reach any node $v$ in the target set $(\bar{V}_{i}^{1}(t)\cup\bar{V}_{j}^{1}(t))/\{u\}$ for the first time. Obviously, node $v$ and node $u$ are adjacent in the previous generation network $SG_3(t-1)$, so the following relationship can be obtained:
\begin{align}\label{3.2}
    T_{u,v}^{3}(t)=\lambda\cdot T_{u,v}^{3}(t-1),~~~\forall u,v\in V(t-1).
\end{align}
Furthermore, the following relationship can be proved:
\begin{align}\label{3.2(1)}
    \sum_{u \in \bar{V}^1(t)}T_{u}^{3}(t)=\lambda\cdot 
    \sum_{u \in {V}(t-1)}T_{u}^{3}(t-1),
\end{align}
where $\bar{V}^1(t)={V}(t-1)$.

Then, for the node set $\hat{V}_{i}^{1}(t)$ inside $X_i(1)$, it is obvious that these nodes are all newly generated nodes in the $t-$th iteration. Starting from node $\forall h \in \hat{V}_{i}^{1}(t)$, the path of walker reaching trap node $2$ can be divided into two sections: (1) starting from node $h$ and reaching $\forall u\in \bar{V}_{i}^{1}(t)$ for the first time; (2) starting from node $u$ and finally reaching trap node $2$. Due to the symmetry of the local region $X_i(1)$, the four nodes in $\bar{V}_{i}^{1}(t)$ will receive walkers starting from the six nodes in $\hat{V}_{i}^{1}(t)$ with equal probability.
Therefore, the following relationship is naturally established:
\begin{align}\label{3.3}
    \sum_{h\in \hat{V}_{i}^{1}(t)}T_{h}^{3}(t)=6\cdot\sigma+ \frac{6}{4}\sum_{u\in \bar{V}_{i}^{1}(t)}T_{u}^{3}(t).
\end{align}

By observing the structure of the network $SG_3(t)$, it can be found that, except for nodes $1$, $2$, and $3$, the degrees of the other nodes are all $6$, that is, except for the three outermost nodes, the remaining nodes are all common nodes of the two smallest regions $X_{i}(0)$ and $X_{j}(0)$ at the same time.
Therefore, based on Eq.(\ref{3.2(1)}) and Eq.(\ref{3.3}), we can decompose the sum of the MCTs of all nodes in the network $SG_3(t)$ as follows:
\begin{align}\label{3.4}
    T_{sum}^{3}(t)=&\sum_{u\in \bar{V}^{1}(t)} T_{u}^{3}(t) +
    \sum_{h\in \hat{V}^{1}(t)} T_{h}^{3}(t)\nonumber\\
    =&\lambda\cdot \sum_{u\in \bar{V}^{1}(t)} T_{u}^{3}(t-1) + 
    6\cdot\sigma\cdot|\Delta^{1}(t)| + 3\cdot \lambda \cdot \sum_{u\in \bar{V}^{1}(t)} T_{u}^{3}(t-1) \nonumber\\
    &- \frac{3}{2}\left(T_{1}^{3}(t)+T_{3}^{3}(t)+T_{4}^{3}(t)\right)
\end{align}
It is easy to get from the initial network $SG3(0)$:
$$
T_{1}^{3}(0)=T_{3}^{3}(0)=T_{4}^{3}(0)=3.
$$
Therefore, $T_{1}^{3}(t)=T_{3}^{3}(t)=T_{4}^{3}(t)=3\cdot\lambda^{t}$. By putting this condition into Eq.(\ref{3.4}), it can be obtained that
\begin{align}\label{3.5}
    T_{sum}^{3}(t)
    =& \lambda\cdot T_{sum}^{3}(t-1) + 
    6\cdot\sigma\cdot|\Delta^{1}(t)| + 3\cdot \lambda \cdot T_{sum}^{3}(t-1) - \frac{27}{2}\cdot \lambda^{t}\nonumber\\
    =& 24\cdot T_{sum}^{3}(t-1) + \frac{9}{2} \cdot 4^{t} - \frac{27}{2}\cdot 6^{t}\nonumber\\
    =& T_{sum}^{3}(0)\cdot 24^{t} + \frac{9}{2} \cdot 4^{t} \cdot\sum_{i=0}^{t-1}6^{i} - \frac{27}{2}\cdot 6^{t} \cdot\sum_{i=0}^{t-1}4^{i}\nonumber\\
    =& \frac{27}{5} \cdot 4^t \cdot 6^t + \frac{9}{2}\cdot 6^t - \frac{9}{10} 4^t.
\end{align}

Consequently, the analytical expression for the MTA on the network $SG3(t)$ with trap node $2$ is:
\begin{align}\label{3.6}
    \langle T^3(t) \rangle
    =\frac{1}{2\cdot4^{t}+1}\left(\frac{27}{5} \cdot 4^t \cdot 6^t + \frac{9}{2}\cdot 6^t - \frac{9}{10} 4^t\right).
\end{align}

\begin{figure}[t]
\centering
\includegraphics[scale=0.7]{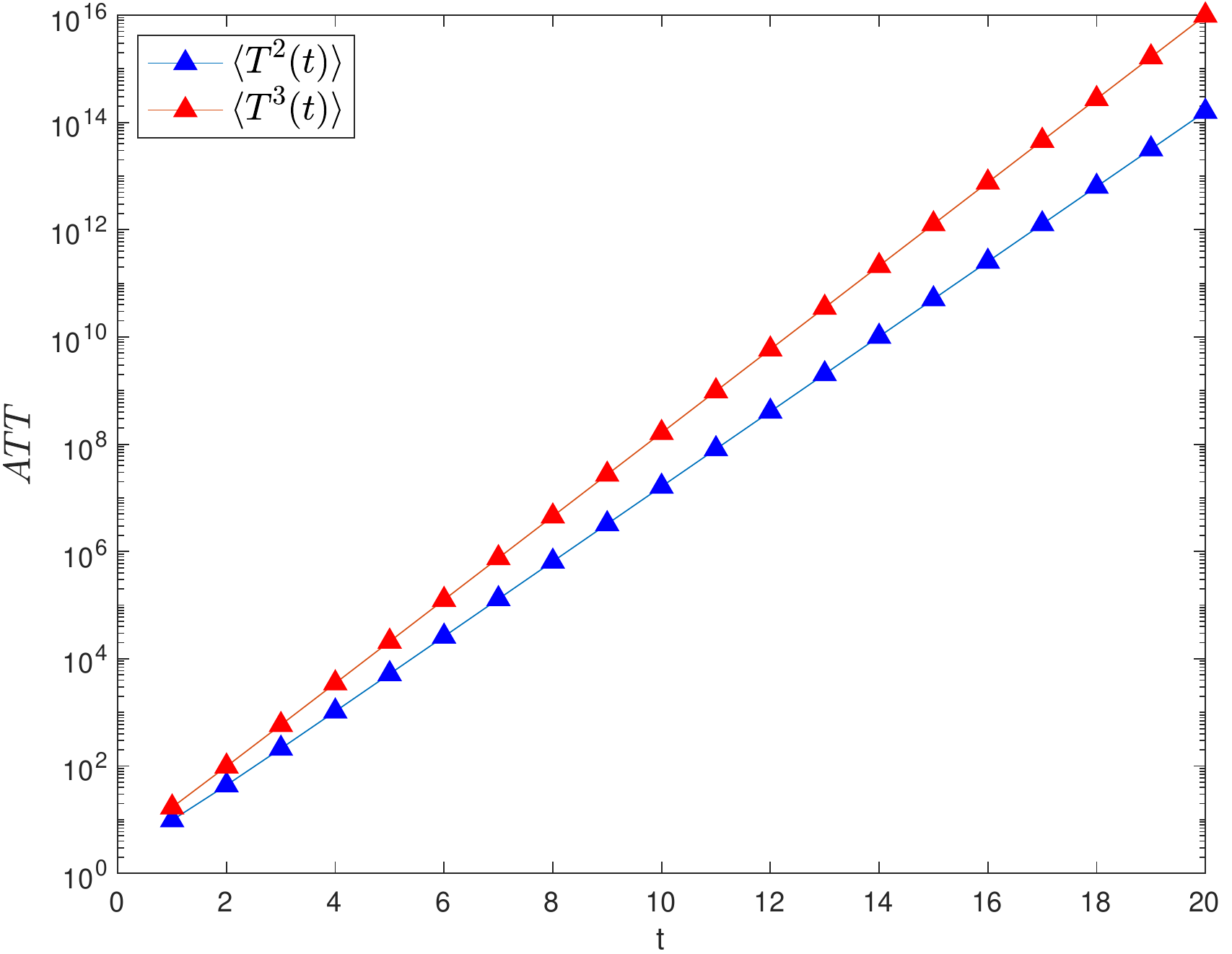}\label{fig3}
\caption{Numerical simulation diagram of $\langle T^3(t) \rangle$ and $\langle T^2(t) \rangle$.}
\end{figure}
In order to compare the incremental relationship of ATT with the number of iterations $t$ on the 2-dimensional and 3-dimensional Sierpinski gasket networks, Fig.(\ref{fig3}) is plotted, where $\langle T^2(t) \rangle$ is the ATT on $SG_2(t)$. Obviously, $\langle T^3(t) \rangle$ and $\langle T^2(t) \rangle$ both increase exponentially with the increase of the number of iterations $t$, and the growth rate of $\langle T^3(t) \rangle$ is faster than that of $\langle T^2(t) \rangle$.

\section{ATT on $ASG(x,y)$}

As mentioned in Section 2, in order to obtain the analytical expression of ATT on $HSG(t,s)$, only the ATT on the auxiliary network $ASG(x,y)$ needs to be solved.

Since the $ASG(x,y)$ is composed of $3^{y}$ local areas $X_i(x)$, the trap node on it is defined as the outermost node $\alpha_2$ of the outermost region $X_1(x)$, which also corresponds to node $2$ in the $HSG(t,s)$.
Similar to the random walk mark in Section 3, in this section we define the MFPT between any two nodes $u$ and $v$ on the network $ASG(x,y)$ as $T^{A}_{u,v}(x,y)$. Then, the MCT of node $u$, the sum of the MCT of all nodes, and the ATT on $ASG(x,y)$ are denoted as $T^{A}_{u}(x,y)$, $T^{A}_{sum}(x,y)$, and $\langle T^{A}(x,y) \rangle$, respectively.

First, we consider the equivalent structure $SG_3(x)$ of the local area $X_i(x)$. Let
$$
T^{3}_{sumA}(x)=\sum_{u\in V(x)/A} T^{3}_{u,A} ~~\textrm{and}~~
T^{3}_{sumB}(x)=\sum_{u\in V(x)/B} T^{3}_{u,B}.
$$
It can be noted that starting from node $u\in V(t)/B$, the path of walkers reaching trap node $2$ can also be divided into two sections in another way: (1) starting from node $u$ to any node in set $B$, (2) if the reached node is not $2$, then starting from that node and finally reach trap node.
Similarly, since the nodes $1$, $2$, $3$, $4$ in the network $SG_3(x)$ are symmetrical, the four nodes in $B$ will receive walkers starting from the nodes in $V(t)/B$ with equal probability. Therefore, the following relationship holds:
$$
T^{3}_{sum}(x)=T^{3}_{sumB}(x)+\frac{3}{4}(|V(x)|-4)\cdot T_{1}^{3}(x) + 3 \cdot T_{1}^{3}(x)
$$
Because the analytical expressions of $T^{3}_{sum}(x)$ and $T_{1}^{3}(x)$ are known, we can obtain:
\begin{align}\label{4.1}
    T^{3}_{sumB}(x)=\frac{9}{10}\cdot4^{x}\cdot6^{x}-\frac{9}{10}\cdot4^{x}.
\end{align}
From Eq.(\ref{4.1}), it is easy to deduce the analytical expression of $T^{3}_{sumA}(x)$ as:
\begin{align}\label{4.2}
    T^{3}_{sumA}(x)&=T^{3}_{sumB}(x) + \frac{1}{4}(|V(x)|-4)\cdot T_{1}^{3}(x) + T_{1}^{3}(x) \nonumber\\
    &= \frac{7}{5}\cdot4^{x}\cdot6^{x}-\frac{9}{10}\cdot4^{x}+ \frac{1}{2}\cdot6^{x}.
\end{align}

Let $C=\{3,4\}$. Then, the MFPT of the walker from node $1$ to the target node set $C$ on $SG_3(t)$ is denoted as $T^{3}_{1,C}(x)$. Because $T^{3}_{1,C}(x)$ is known and the walker starting from node $1$ will eventually reach any node in the set $A$ with the same probability, it can be obtained that
\begin{align}\label{4.3}
    T^{3}_{1,C}(x)=T_{1,A}(x) + \frac{1}{3}T^{3}_{1,C}(x)
    \Rightarrow  T^{3}_{1,C}(x)=\frac{3}{2}\cdot \lambda^{x}.
\end{align}

Based on the above quantitative analysis of the property of random walks on network $SG_3(x)$, we can explore the MCTs of nodes in sets $\tilde{V}^A(x,y)$ and $\dot{V}^A(x,y)$ respectively.

In the auxiliary network $ASG(x,y)$, each self-similar region $X_{i}(x)$ is equivalent to the network $SG_3(x)$, and the node $u\in \tilde{V}^A(x,y)$ is at most a common node of two regions. If $u\in \tilde{V}^A(x,y)$ is the common node of $X_{i}(x)$ and $X_{j}(x)$, then $X_{i}(x)$ and $X_{j}(x)$ are symmetrical with respect to node $u$. 
Therefore, from Eq.(\ref{4.3}), we can see that the mean time for a walker to depart from node $u$ and reach any node $v\in (\tilde{V}^A_{i}(x,y)\cup \tilde{V}^A_{j}(x,y))/\{u\}$ for the first time is $T^{3}_{1,C}(x)$. In addition, the nodes $u$ and $v$ are adjacent in the two-dimensional network $SG_2(y)$. Accordingly, the following equation is naturally proven:
\begin{align}\label{4.4}
    T^{A}_{u,v}(x,y)=\frac{3}{2}\cdot \lambda^{x}\cdot T^{2}_{u,v}(y)~~~\forall u,v \in \tilde{V}^A(x,y),
\end{align}
where $T^{2}_{u,v}(y)$ is the MFPT from node $u$ to node $v$ in $SG_2(y)$. Naturally, it can be obtained that:
\begin{align}\label{4.4(1)}
    \sum_{u \in \tilde{V}^A(x,y)} T^{A}_{u}(x,y)
    =\frac{3}{2}\cdot \lambda^{x}\cdot \sum_{u \in \tilde{V}^A(x,y)} T^{2}_{u}(y)
    =\frac{3}{2}\cdot \lambda^{x}\cdot T^{2}_{sum}(y)
\end{align}

In addition, the MFPT from node $h\in\dot{V}^{A}_{i}(x,y)$ to the trap node is analyzed by way of path segmentation. The path of walker starting from node $h$ and finally reaching the trap node can be divided into two independent segments: (1) starting from node $h$, it first reaches $v\in \tilde{V}^A_{i}(x,y)$; (2) starting from node $v$ and finally reaching the trap node. Due to the symmetry of nodes $2_i$, $3_i$, and $4_i$ in $X_i(x)$, these three nodes will receive all walkers starting from the node set $\dot{V}^{A}_{i}(x,y)$ with the same probability. Therefore, based on Eq.(\ref{4.2}), the following relationship can be obtained:
\begin{align}\label{4.5}
    \sum_{h\in \dot{V}^{A}_{i}(x,y)} T_{h}^{A}(x,y) 
    &= T^{3}_{sumA}(x) + \frac{|V(t)-3|}{3} (T^{A}_{2_i}(x,y)+T^{A}_{3_i}(x,y)+T^{A}_{4_i}(x,y))\nonumber\\
    &= T^{3}_{sumA}(x) + \frac{|V(t)-3|}{2} \cdot \lambda^{x}
    \cdot (T^{2}_{2_i}(y)+T^{2}_{3_i}(y)+T^{2}_{4_i}(y))
\end{align}

\begin{figure}[t]
\centering
\includegraphics[scale=0.7]{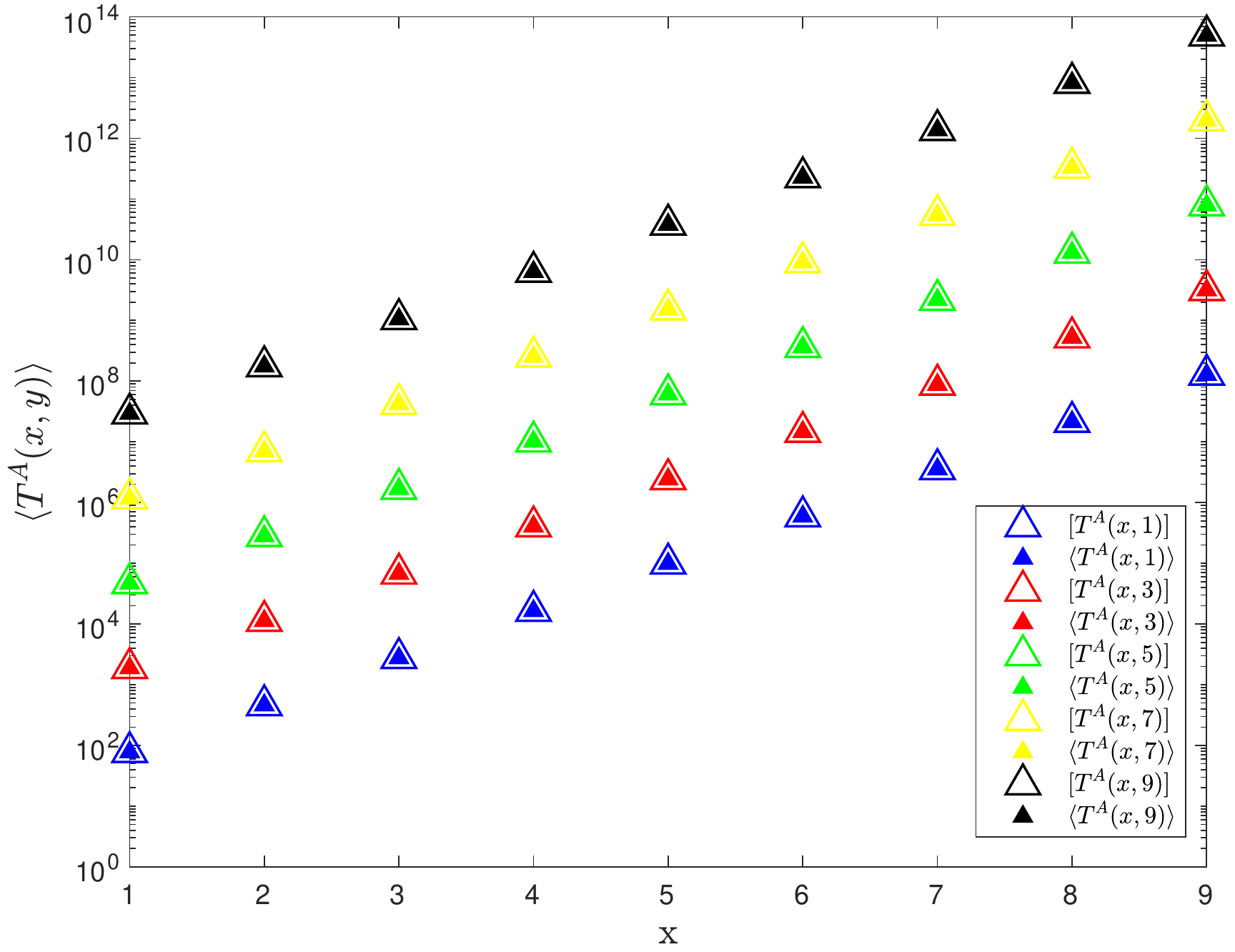}\label{fig4}
\caption{Numerical simulation diagram of $\langle T^{A}(x,y) \rangle$.}
\end{figure}

By observing the structure of the network $ASG(x,y)$, it can be known that, except for the three outermost nodes in $\tilde{V}^A(x,y)$, the remaining nodes are all common nodes in two local self-similar regions. Hence, based on Eq.(\ref{4.4(1)}) and Eq.(\ref{4.5}), the sum of the MCTs of all nodes on the network $ASG(x,y)$ can be expressed as follows:
\begin{align}\label{4.6}
    T^{A}_{sum}(x,y)
    =&\sum_{u \in \tilde{V}^A(x,y)} T^{A}_{u}(x,y) + 
     \sum_{i=1}^{3^y}\sum_{h \in \dot{V}^{A}_{i}(x,y)}  T^{A}_{h}(x,y) \nonumber\\
    =& \frac{3}{2}\cdot \lambda^{x}\cdot T^{2}_{sum}(y) + 
     \sum_{i=1}^{3^y} \Big[T^{3}_{sumA}(x) + \frac{|V(t)|-3}{2} \cdot \lambda^{x}\nonumber\\
    &\cdot (T^{2}_{2_i}(y)+T^{2}_{3_i}(y)+T^{2}_{4_i}(y))\Big]\nonumber\\
    =& \frac{3}{2}\cdot \lambda^{x}\cdot T^{2}_{sum}(y) +
     3^{y}\cdot T^{3}_{sumA}(x) + \frac{|V(t)|-3}{2} \cdot \lambda^{x}\cdot \nonumber\\
    & \Big[2T^{2}_{sum}(y) - T^{2}_{\alpha_2}(y) - T^{2}_{\alpha_3}(y) \Big]
\end{align}
From previous research\cite{zhang2021mean}, we know: $T^{2}_{\alpha_2}(y)=T^{2}_{\alpha_3}(y)=2\cdot 5^{y}$, so Eq.(\ref{4.6}) can be further expressed as:
\begin{align}\label{4.7}
    T^{A}_{sum}(x,y)
    =& \big[|V(t)|-\frac{3}{2}\big]\lambda^{x}\cdot T^{2}_{sum}(y) +3^{y}\cdot T^{3}_{sumA}(x) - (|V(t)|-3) \lambda^{x}\cdot T^{2}_{\alpha_2}(y)\nonumber\\
    =&5\cdot 15^y\cdot 24^x + \frac{2}{5} \cdot 3^y \cdot 24^x + \frac{5}{4}\cdot 15^y \cdot 6^x + 3\cdot 5^y \cdot 6^x \nonumber\\
    &+ \frac{1}{4}\cdot 3^{y} \cdot 6^{x}  - \frac{9}{10} \cdot 3^x \cdot 4^{y}.
\end{align}

Finally, from the definition of ATT, the analytical expression of the ATT on the network $ASG(x,y)$ with trap node $\alpha_2$ is as follows:
\begin{align}\label{4.8}
    \langle T^{A}(x,y) \rangle 
    =&\frac{T^{A}_{sum}(x,y)}{|V^{A}(x,y)|-1}\nonumber\\
    =&\frac{1}{40\cdot 3^y\cdot 4^x + 10 \cdot 3^y + 10}\Big[100\cdot 15^y\cdot 24^x + 8 \cdot 3^y \cdot 24^x  \nonumber\\
    & + 25\cdot 15^y \cdot 6^x+ 60\cdot 5^y \cdot 6^x + 5\cdot 3^{y} \cdot 6^{x}  - 18 \cdot 3^y \cdot 4^{x}\Big]
\end{align}

In order to verify the accuracy of the calculation conclusions in this section, numerical simulation Fig.(\ref{fig4}) is illustrated, where $[T^{A}(x,y)]$ is the numerical result obtained by the matrix algorithm of the computer. When the independent variables $x$ and $y$ take different values, the numerical results obtained by the computer are consistent with the results obtained by the analytical expression of Eq.(\ref{4.8}), which also proves the accuracy of the above conclusions.

\section{ATT on $HSG(t,s)$ and simulations}

\begin{figure}[t]
\centering
\includegraphics[scale=0.7]{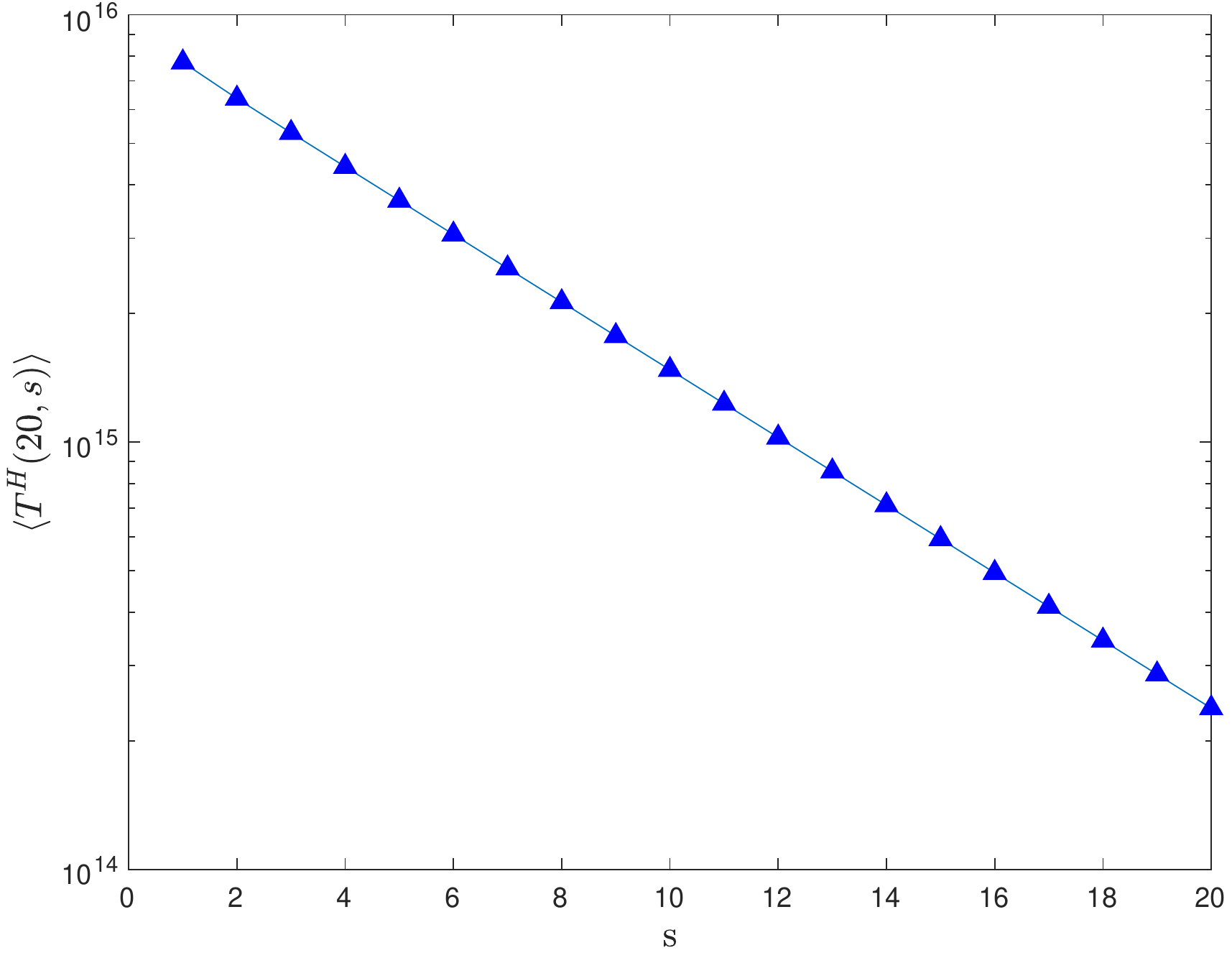}\label{fig5}
\caption{When $t=20$, the change trend graph of $\langle T^{H}(t,s) \rangle$ with respect to the segmentation coefficient $s$.}
\end{figure}

In the previous section, the ATT on the auxiliary network $ASG(x,y)$ with the trap node $\alpha_2$ has been solved.
In addition, as described in Section 2, the horizontally segmented 3 dimensional Sierpinski gasket network $HSG(t,s)$ can be constructed by the auxiliary network $ASG(x,y)$, that is: $HSG(t,s)=ASG(t-s,s)$, so the analytical expression of the ATT on the $HSG(t,s)$ with the trap node $2$, denoted as $\langle T^{H}(t,s) \rangle$, is as follows:
\begin{align}\label{5.1}
    \langle T^{H}(t,s) \rangle 
    =&\langle T^{A}(t-s,s) \rangle\nonumber\\
    =&\frac{1}{40\cdot 18^s\cdot 4^t + 10 \cdot 72^s + 10\cdot 24^s}\Big[100\cdot 15^s\cdot 24^t + 8 \cdot 3^s \cdot 24^t  \nonumber\\
    & + 25\cdot 60^s \cdot 6^t+ 60\cdot 20^s \cdot 6^t + 5\cdot 12^{s} \cdot 6^{t}  - 18 \cdot 18^s \cdot 4^{t}\Big]
\end{align}

In order to explore the changing law of $\langle T^{H}(t,s) \rangle$ with the segmentation coefficient $s$, Fig.(\ref{fig5}) is drawn, in which the number of iterations $t$ is fixed to $20$. Obviously, as the segmentation coefficient $s$ increases, the ATT on the network $HSG(t,s)$ will decrease exponentially. However, since $s\in [0,t]$, there is a lower bound $\langle T^{H}(t,t) \rangle$, and it's analytical expression is:
\begin{align}\label{5.2}
    \langle T^{H}(t,t) \rangle 
    =&\frac{1}{10\cdot 3^t + 2}\Big[25\cdot 15^t  + 12\cdot 5^t   - 3^t\Big]\sim \langle T^{2}(t) \rangle
\end{align}
Combined with the difference between $\langle T^{2}(t) \rangle$ and $\langle T^{3}(t) \rangle$ regarding the increasing speed of the number of iterations $t$, the following judgments can be obtained: This segmentation process realizes the transformation of ATT from the 3-dimensional network $SG3$ to the two-dimensional network $SG2$, that is, the larger the segmentation coefficient $s$, the ATT on the $HSG(t,s)$ is indeed close to the ATT of $SG_2(t)$.

\begin{figure}[t]
\centering
\includegraphics[scale=0.7]{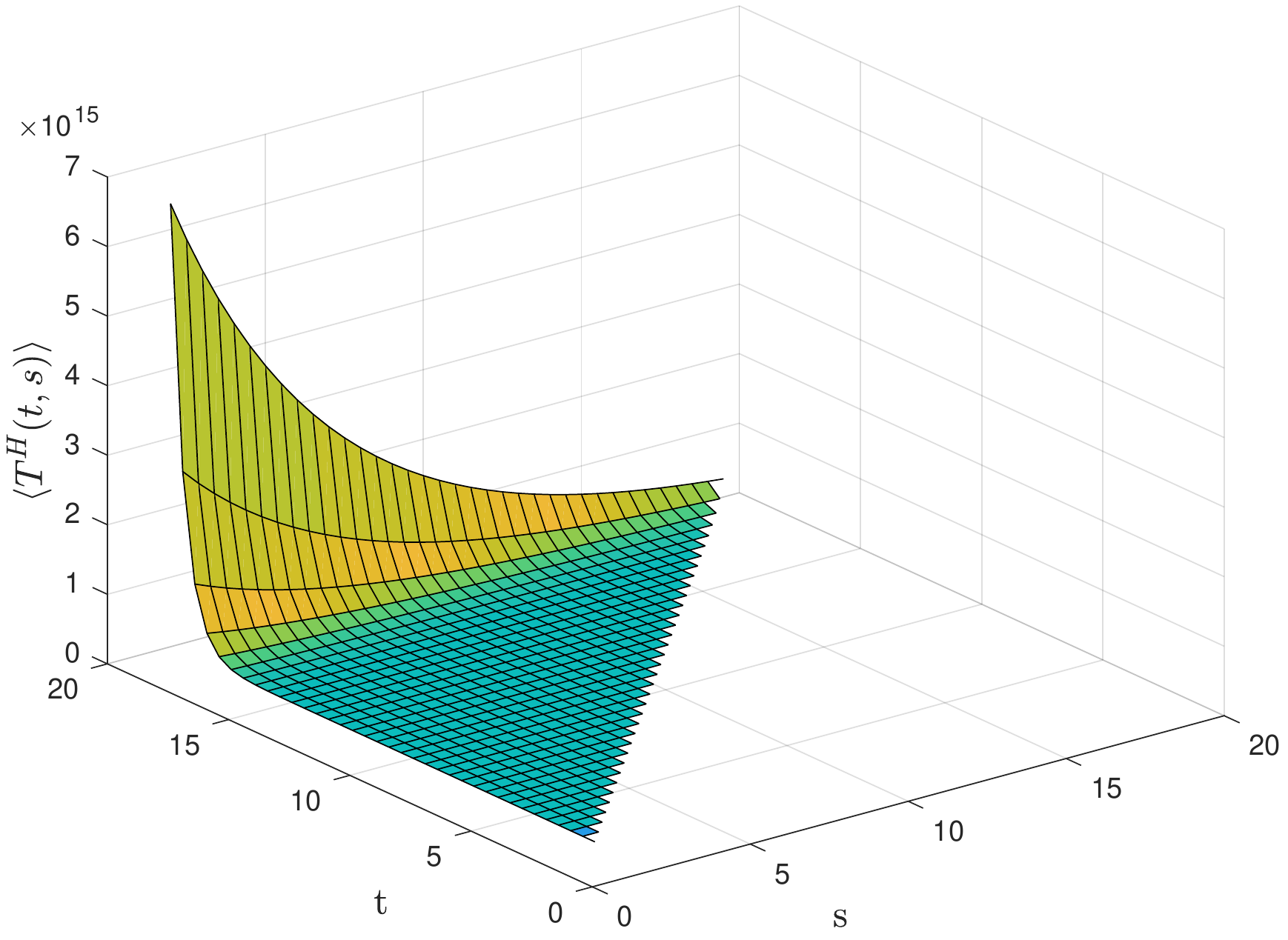}\label{fig6}
\caption{Numerical simulation diagram of $\langle T^{H}(t,s) \rangle$ with respect to variables $t$ and $s$.}
\end{figure}

In order to explore the change trend of ATT on $HSG(t,s)$ with the number of iterations $t$ when the segmentation coefficient $s$ is fixed, a three-dimensional numerical simulation diagram of $\langle T^{H}(t,s) \rangle$ with respect to variables $t$ and $s$ is drawn, as shown in Fig.(\ref{fig6}). Obviously, even when the segmentation coefficient $s$ is different, $\langle T^{H}(t,s) \rangle$ will still increase exponentially at the same speed as the number of iterations $t$ increases.

In fact, because the 2-dimensional network $SG_2(t)$ and the 3-dimensional network $SG_3(t)$ both have independent self-similar structures, the segmentation process constructed in this paper actually realizes the gradual transformation from $SG_3(t)$ to $SG_2(t)$, during which the self-similar structure of $SG_3(t)$ is gradually lost leading role, and the self-similar structure of $SG_3(t)$ gradually dominates. 
Such dynamic structural characteristics are reflected in ATT as $\langle T^{H}(t,s) \rangle$ gradually decreases with the increase of the segmentation coefficient $s$, and tends to $\langle T^{2}(t) \rangle$. 
But when $s$ is fixed, as $t$ increases, the self-similar structure of $SG_3(t)$ will gradually return to the dominant position, that is, as the number of iterations $t$ increases, $\langle T^{H}(t,s) \rangle$ will increase exponentially at the same speed as $\langle T^{3}(t) \rangle$.
The above conclusions together prove that in a network with multiple local self-similar structures, each self-similar structure will have an impact on the properties of random walks, but the dominant structural features will have a greater impact.

\section{Conclusion}

In this article, a horizontally segmented 3 dimensional Sierpinski gasket network $HSG(t,s)$ is constructed and the analytical expressions for ATT of specific trap nodes are solved on it. 
First, in the second section, the construction methods and structural characteristics of the 2-dimensional Sierpinski gasket network $SG_2(t)$, the 3-dimensional Sierpinski gasket network $SG_3(t)$, and the horizontally segmented 3 dimensional Sierpinski gasket network $HSG(t,s)$ are explained. 
Secondly, in the third section, based on the analysis of the walking path of the walker to the target node, the analytical expression of ATT on the network $SG_3(t)$ with trap node $2$ is solved.
Then, using the ATT on the networks $SG_3(t)$ and $SG_2(t)$, the analytical expression of the ATT on the auxiliary network $ASG(x,y)$ with the trap node $\alpha_2$ is obtained, where the structure of the network $ASG(x,y)$ is closely related to the network $HSG(t,s)$.
Finally, based on the structural relevance of the network $ASG(x,y)$ and the network $HSG(t,s)$, by setting specific coefficients, the analytical expression of the ATT on the $HSG(t,s)$  with the trap node $2$ is obtained. 
Through the analysis of ATT on $HSG(t,s)$, we can finally conclude that the dominant self-similar structure in the network will have a greater impact on the property of random walks.

\bibliographystyle{unsrt}

\bibliography{ref}

\end{document}